\documentclass[preprint,12pt]{elsarticle}

\journal{ArXiv}

\begin{document}

\begin{frontmatter}

\title{"Grating Echo" instrument for small-angle neutron scattering investigations}

\author[1]{Y.O.Chetverikov}
\author[2]{S.A.Ivanov}

\address[1]{PNPI, NRC KI, St. Petersburg-Gatchina}
\address[2]{ITMO University, St. Petersburg}

\begin{abstract}

The concept of the instrument with absorption gratings into small-angle neutron scattering measuring is discussed. The instrument action consists of passing direct and scattered radiation through a system of 3 absorption gratings, consisting of periodic transparent and absorbing strips. The instrument operates in a mode of geometric optics. The principle of operation of the instrument is demonstrated on a prototype created for optical radiation. For the instrument equations of the radiation transmission and the display of the sample, scattering function has been derived. The necessary parameters of neutron absorption gratings and design limitations of the installation are given. By the concept of "grating echo instrument for neutron scattering", the expected characteristics and experimental possibilities are presented.

\end{abstract}


\end{frontmatter}

\newpage
\section{Introduction}
\label{Intro}

Based on absorption periodic gratings tools for neutron and x-ray small-angle scattering investigations have been known for a relatively long time \cite{Ioffe85} \cite{Lebedev93},  however, over the past decade, there has been a rapid increase in activity in the development of such systems \cite{Strobl14}. The use of an interferometer of absorption gratings makes it possible to extract the signal of small-angle scattering (SAS) contribution of an object while simultaneously ensuring spatial resolution of the scattering pattern. This opens up new possibilities for studying spatial inhomogeneities of the internal structure of objects using small-angle scattering.

The use of a grating interferometer not only as an effective tool for small-angle introscopy, but also as a fast technique for the scattering function measuring are demonstrates in this paper. The use of the grating interferometry technique will increase the experimental capabilities of existing small-angle scattering installations by increasing the measuring range, increase the efficiency of SAS studies by reducing the measurement time in existing and promising instruments.

\section{The optical prototype of the neutron instrument and obtained results}
\label{Layout}

\begin{figure}[htb]
\includegraphics[scale=0.4]{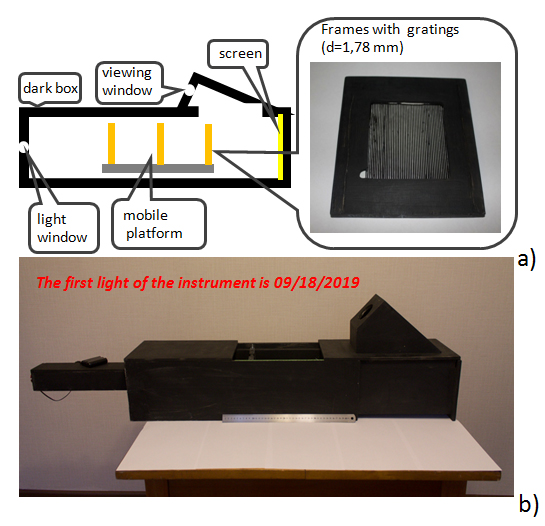}
\centering
\caption{ The prototype the instrument on the absorption gratings: a) sketch of the prototype ; b) photo of the prototype
}
\label{Fig1}
\end{figure}

To identify the operability of the general scheme for constructing the instrument, a prototype that works with optical radiation was created (Fig.\ref{Fig1}). The model of the interferometer is located in a dark box, which reduces the light background of external radiation sources and designed as the support-ing construction for the elements of the interferometer. The box has a light window for the radiation source (a white LED with a power of the order of 3 Wt was used) and an observation window for recording scattered radiation (as a recorder Canon 1100D camera with a 28-55 mm lens was used).

\begin{figure}[h]
\includegraphics[scale=0.45]{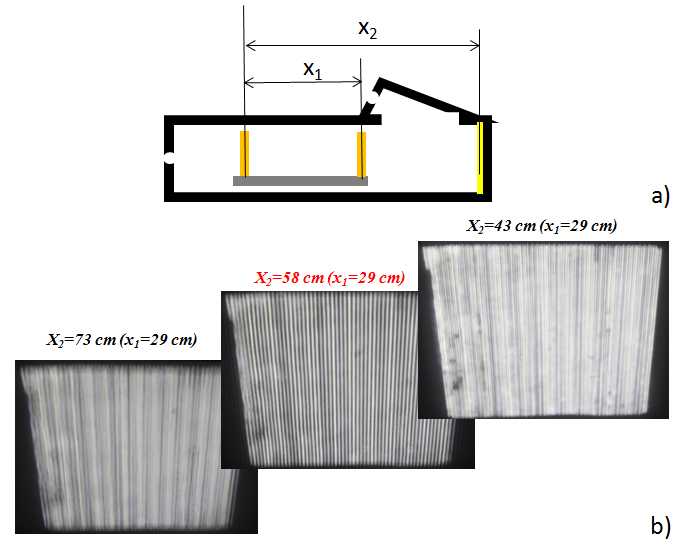}
\centering
\caption{ Demonstration of the effect of light echo: a) sketch of the experiment; b) screenshots}
\label{Fig2}
\end{figure}

Inside the box, there is a movable platform with gratings printed on the glass and a projection screen for displaying an interference pattern.
A light from a source with a cross-section greater than the period of the gratings is passing through two identical gratings of periodic strips placed at a distance $x_1$ creates the following light picture. On a screen located at a distance $x_2=x_1$ after the second grating, alternating bright and dark stripes of the same period as the stripes themselves will be observed gratings. At a shorter ($x_2<x_1$) or longer ($x_2> x_1$) distance between the screen and the gratings, blurring of the bright and dark stripes will be observed (Fig.\ref{Fig2}).

\begin{figure}[h]
\includegraphics[scale=0.5]{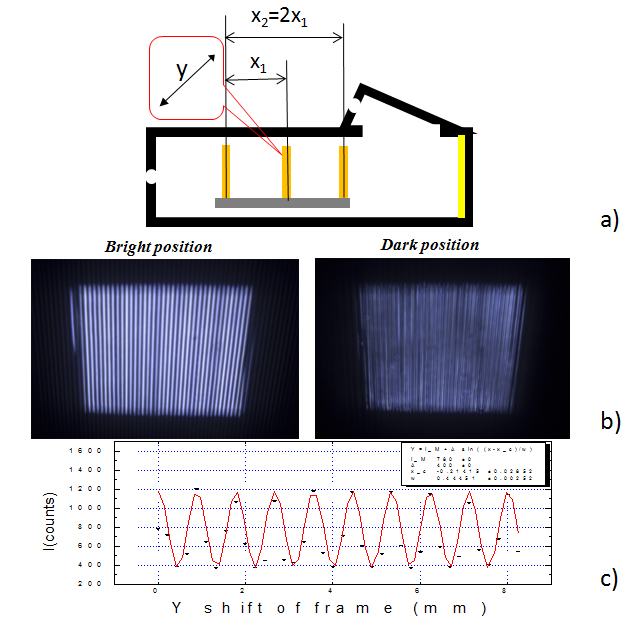}
\centering
\caption{ Bright and dark positions of the interferometer:
a) sketch of the experiment; b) screenshots of bright and dark positions ; c) dependence of the intensity on the screen on the transverse displacement (along the $y$ axis) of one of the gratings
}
\label{Fig3}
\end{figure}

Putting in the position of the screen (at a distance $x_2=x_1$ from the second grating) the same grating as the previous two, and shifting the screen further in the direction of radiation we get either a dark screen (if the black stripes of the grating coincide with the bright stripes of the image) or a bright screen (if the transparent stripes of the grating coincide with the bright stripes of the image; Fig.\ref{Fig3}, a). Moving one of the gratings across the beam in the y-direction, perpendicular to the direction of the strips, we get a picture of periodic oscillations of light and dark screen projections with a period $d_O = d_{G2}$, where $d_G$ is the period of the grating (Fig.\ref{Fig3}, c).

The observed contrast between the light and dark positions of the gratings will be a measurable quantity and can be written as:

\begin{equation}
P_{GE}=(I_L-I_D)/(I_L+I_D)
\label{eq:contrast}
\end{equation}
where $I_L$, $I_D$- intensities of bright and dark positions respectively.

\begin{figure}[h]
\includegraphics[scale=0.5]{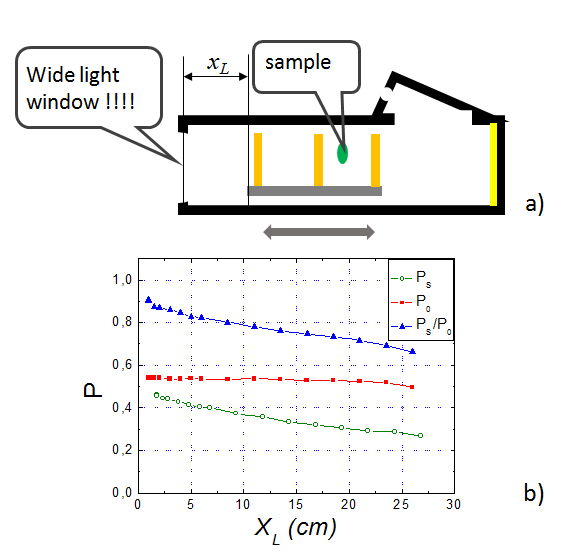}
\centering
\caption{Contrast between bright and dark positions: a) sketch of the experiment; b) The dependence of contrast on the $x$- position of the platform with gratings: measurements without a sample ($P_0$) and with a sample of ground coffee ($P_S$); c) normalized values ($P_S/P_0$) }
\label{Fig4}
\end{figure}

A key feature of the GE SAS method compared to traditional SAS meth-ods is that the measured contrast value P is practically independent of the area and divergence of the radiation incident on the sample, and it also weakly depends on the position $x_L$ characterizing the relative position of the light source-screen system relative to a moving platform with gratings (scheme 4 a, red dots on the graph of Fig. \ref{Fig4}, b).

Such a system of three gratings can be used to measure small-angle scattering on the sample: the sample is placed between the gratings, and the ratio of intensities in the dark and bright positions varies depending on a complex of parameters such as  - the scattering function of the sample;-grating period; - radiation wavelength (in the measurements carried out on the prototype, the intensity of only the green channel /$\lambda$ = 550 nm/ of the camera was analyzed ); mutual arrangement of gratings; - the position of the sample relative to a movable platform with gratings.

Scanning occurs when one of the parameters changes. In this prototype, scanning of the scattering function of the sample is carried out by changing the position of the moving platform (blue and green dots on the graph of Fig. \ref{Fig4}, b.).

\section{Description of "Grating Echo"}
\label{Theory}

\begin{figure}[h]
\includegraphics[scale=0.5]{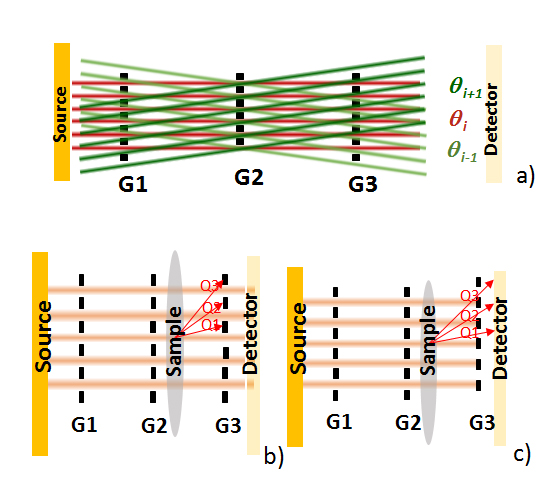}
\centering
\caption{Transmission and features of the visualization of the scattering: a) GE SAS instrument as a set of collimated beams; b) visibility of the scattering in a bright position; c) visibility of the scattering in a dark position}
\label{Fig5}
\end{figure}

The optical scheme of the device can be represented by a set of n collimated beams (n is the number of slits in the grating), each of which has the same divergence (determined by the size of the slits and the distance between adjacent grating) but its own angular spatial orientation (Fig. \ref{Fig5},a.).

Each grating absorbs radiation by more than 2 times; therefore, the intensity at the detector decreases by $T=T_{G1}*T_{G2}*T_{G3}$ times, where $T_{Gi}$ is the transmission coefficient of the grating. The radiation transmission of the GE SAS tool compared to the usual SAS instrument:

\begin{equation}
T_{GE}/T_{SAS}=n_{min}/T_{G1}T_{G2}T_{G3}
\label{eq:transmission}
\end{equation}

$n_{min}$ - the number of stripes on the smallest grating.

To describe the scattering visualization obtained by the GE SAS instrument, two limiting tuning cases should be considered:

a)	GE SAS instrument in the bright position: all the radiation falls on the detector, except for the scattered in the strips of the third grating, then by the scattering scheme shown in Fig. \ref{Fig5} b. the intensity on the screen will be:
\begin{equation}
I_L = I_0 –I_0 (S(Q_1) + S(Q_2) + S(Q_3) + ...)
\label{eq:BrightInt}
\end{equation}
where $Q_i = iQ_1$; $I_0$ - intensity on the screen in the absence of a sample; $S(Q)$ is the scattering function of the sample.

b)	GE SAS in the dark position: only scattered radiation which passes the absorption bands of the third grating enters the screen:
\begin{equation}
I_D = I_0 (S (Q_1) + S (Q_2) + SQ_3 + ...)
\label{eq:DarkInt}
\end{equation}

Introducing the contrast from Eq. \ref{eq:contrast} as a "polarization of grating echo" and defining "grating echo length" as:
\begin{equation}
\delta_{GE} = 2 \pi/Q_1
\label{eq:zGE}
\end{equation}
we make the whole formalism of the SESANS \cite{Rekveldt05} method relevant. Then the "GE polarization" of the scattered signal will be described as:
\begin{equation}
<P_{GE}> = 1/(s k_0)\int S_y(Q)cos(Q_y \delta)dQ;
<P_{GE}>=G_{GE}(\delta)
\label{eq:GGE}
\end{equation}

where $G_{GE}$ is the "GE correlation function" of the sample, which is the Fourier transform of the projection $ S_y(Q)$ of the scattering function of the sample $S(Q)$.

\begin{section}{Measurements examples on the optical prototype instrument with absorbtion gratings}
\label{Exemples}

To test the health of the prototype, measurements of improvised materials were carried out. An example of measuring powder from ground coffee is shown in Fig. \ref{Fig6}.
\begin{figure}[h]
\includegraphics[scale=0.55]{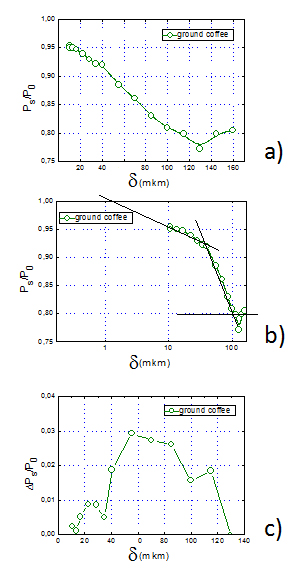}
\centering
\caption{The scattering on a ground coffee: graphs of the measurement in a linear (a) and non-linear (b) scale; variation of image contrast depending on the "scattering length" $\delta$ }
\label{Fig6}
\end{figure}
Ground coffee was applied to an adhesive tape, which was installed in a light window located across the light beam. The scattering on the adhesive tape itself was measured separately, turned out to be relatively small and the result was normalized to the value of this scattering.

Graphs \ref{Fig6} a and \ref{Fig6} b shows the scattering curve of ground coffee as a function of the reducing contrast of the display of gratings  $P_S / P_0$ depending on the "echo length" - the coordinate of the transformation of the position of the sample relative to the platform with gratings into the length $\delta$ (the reciprocal of the characteristic transmitted momentum $Q_1$, see Eq \ref{eq:zGE}). On the linear scale (Fig.\ref{Fig6} a.), the bends of the scattering curve are smooth, it is difficult to distinguish the characteristic scattering scales, while on the non-linear scale (Fig.\ref{Fig6} b.) the bends of the scattering curve are sharp, and the bending sites are visible. From the non- linear curve can be distinguished two scattering regimes with their own slope of the curve; the change of regimes is visible as a sharp change in the slope at $\delta$=40 mkm.

\begin{table}[hp]
\caption{Measurement examples on the optical instrument with absorption gratings}
\includegraphics[scale=0.4]{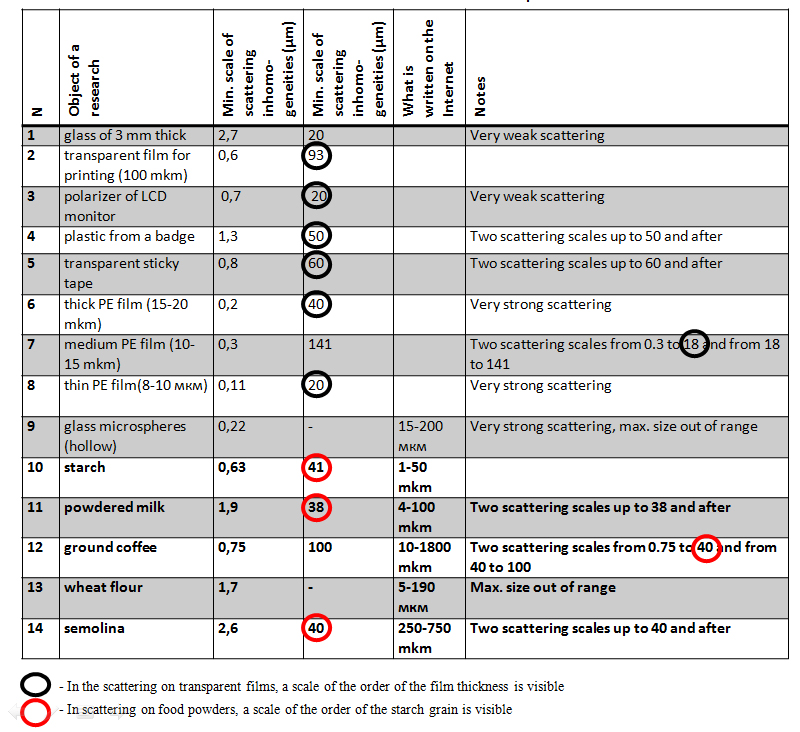}
\centering
\label{Tab1}
\end{table}

Thus, three characteristic scattering particle scales can be distinguished: the minimum scattering particle scale is the interpolated point at which the decrease in $P_S / P_0$ begins; the scale of the scattering regime changes is, as described above, the point on the curve with $\delta$ = 40 mkm; the maximum scattering particle scale is the bending point in a curve where the curve goes into a horizontal line. Thus, a change in the presented contrast $P_S / P_0$ depending on the "echo length " carries information on the particle size distribution (Fig.\ref{Fig6} c.).

Table \ref{Tab1} shows the results of some measurements of optically transparent glass samples and plastic films (items 1-8) and powder samples (items 9-14). The repeatability of the results is at 3$\% $. In the case of weak scattering (contrast $P_S / P_0$ greater than 0.2), the line shape of the scattering curve does not change with a change in the number of samples (the number of layers or the amount of powder). In the film's measurements, a scale of the order of the thickness of the film is clearly noticeable, in the measurement of food powders, a scale of the order of the size of the grains of starch is noticeable too.

It can be seen from the above studies that to increase the information content of measurements, it is necessary to increase the measuring range of the setup, improve the contrast of the gratings, and prepare samples in better quality to avoid too strong or too weak scattering.
\end{section}

\begin{section}{Design features of the implementation of the neutron "grating echo" method}
\label{Features}

To implement the method of "neutron grating echo" on research sources of thermal neutrons, it is necessary to develop effective absorption gratings with a period of the order of 1-20 mkm. The choice of the period of the gratings is due to several circumstances. According to equation \ref{eq:transmission}, to ensure maximum transmission, it is necessary to ensure the maximum number of strips in the grating, which means the minimum possible period of the strips. Since GESANS operates in geometric optics, the minimum period of the gratings is limited by the diffraction limit. When approaching the diffraction limit, the bright and dark strips of the display of the gratings are blurred. Minimum achievable angle in a mode of geometric optics is $\psi=1.22 \lambda/d_S$. Knowing the width of the slit ($d_S$) and the minimum angle  ($\psi$), we can calculate the maximum possible distance between the gratings:
\begin{equation}
L_{MAX}=d_S/sin(\psi )
\label{eq:transmission}
\end{equation}

The diffraction limit also determines the maximum sizes of scattered inhomogeneities which could be investigated by the GESANS instrument as $D=d_S$.

Another important circumstance limiting the radiation transmission of the device is the projectively decreasing of the contrast (Fig. \ref{Fig7}). On the surface of the third grating, the display of strips is projectively blurred, which leads to a decrease in contrast. The maximum contrast is achieved for gratings with a ratio of the widths of transparent and absorption strips as $ r = 1/3$. The transmittance of such gratings is $1/4$; therefore, to achieve the efficiency of the conventional SAS instrument, the diverging radiation of each slit from the first or second gratings must provide illumination of more than 64 slits of adjacent gratings. Then the ratio $T_{GE}/T_{SAS}$ from equation (Eq. \ref{eq:transmission}) will be more than one.

\begin{figure}[htb]
\includegraphics[scale=0.5]{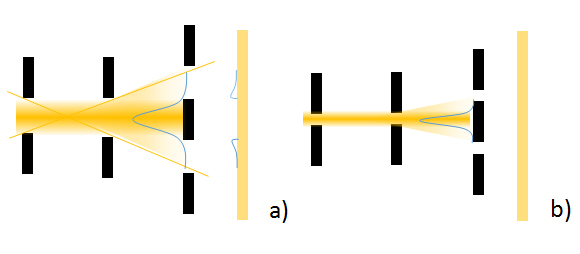}
\centering
\caption{Contrast and transparency of absorption gratings: a) comparatively low contrast of gratings with $1/2$ transparency; b) comparatively high contrast of gratings with 1/4 transparency }
\label{Fig7}
\end{figure}
\begin{figure}[htb]
\includegraphics[scale=0.4]{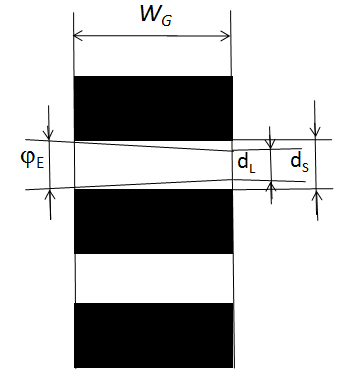}
\centering
\caption{Aperture of neutron absorption gratings: $\varphi_E$ - angular aperture; $W_G$ - grating thickness; $d_S$ - gap width; $d_L$ - light bandwidth }
\label{Fig8}
\end{figure}
\begin{table}[htb]
\caption{Materials and effective apertures of neutron gratings}
\includegraphics[scale=0.4]{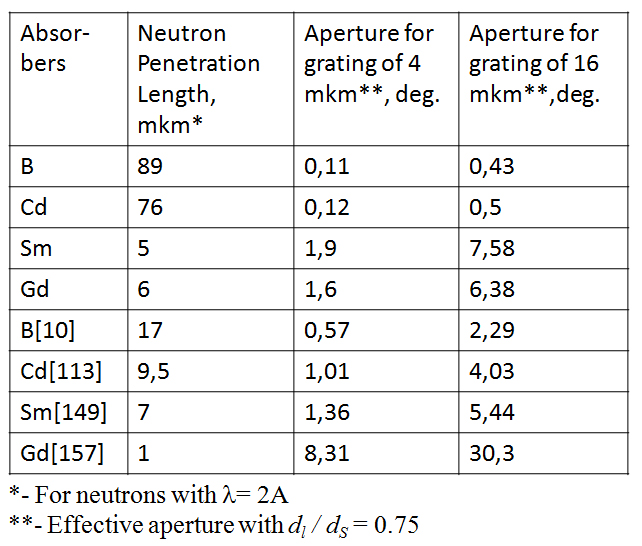}
\centering
\label{Tab2}
\end{table}

The need to ensure a wide angular aperture with a slit size of several microns imposes limitations both on the choice of the absorbent material of the gratings and on the technology for their production. The most effectiveabsorbent materials have a thermal neutron penetration depth greater than a micrometer. This means that the depth of the absorption layer of the strips must be of the order of or greater than the width of the slit. Such a geometry limits the angular aperture of the grating, making it ineffective to use it for neutron beams with large angular divergence (Fig. \ref{Fig8}). Table \ref{Tab2} shows the comparative characteristics of the absorption materials for the manufacture of the gratings.
\end{section}

\begin{section}{The Development Prospects and Possibilities of Using the Method}
\label{Prospects}
The "grating echo" module can be used as an additional option of existing small-angle scattering instruments (Fig. \ref{Fig9}). Having a compact size of less than 1.5 meters along the beam, a platform with three gratings can be placed in the region of the sample: between the collimation tube and PSD detector. Using the module will allow us to increase the measuring range of the instrument, making it possible to study inhomogeneities with a scale of several microns (existing SANS instruments can distinguish inhomogeneities of the order of tenths of a micron). Also, the "grating echo" module can use the entire cross-section of the neutron guide. This feature will allow either to obtain the spatial resolution of scattering of a large sample (with a cross-section of the order of the neutron guide cross-section) or to simultaneously study several samples spaced along the beam cross-section.

\begin{figure}[htb]
\includegraphics[scale=0.5]{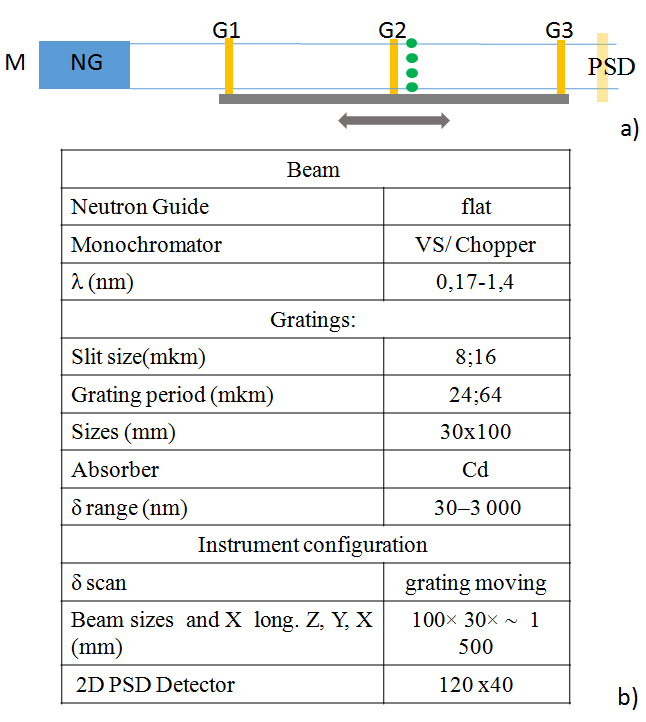}
\centering
\caption{Sketch (a) and parameters (b) of the interferometer module for a SANS instrument }
\label{Fig9}
\end{figure}

\end{section}

\begin{section}{Conclusion}
\label{Conclusion}
\begin{enumerate}
\item The concept of using an interferometer of absorption gratings as an "echo instrument" for small-angle scattering studies is presented;
\item the prototype that clearly demonstrates the operability of the proposed approach for optical radiation was created;
\item successful measurements of scattering on samples using the prototype of the instrument were done;
\item in the measurements advantages and disadvantages of the used prototype and the presented concept as a whole were revealed;
\item the project of neutron instrument that implements the concept of an interferometer was proposed.
\end{enumerate}

\end{section}


\begin{thebibliography}{3}
\bibitem{Ioffe85}
Ioffe, A. I., Zabiyakin, V. S., and Drabkin, G. M. (1985). Test of a diffraction grating neutron interferometer. Physics Letters A, 111(7), 373-375.
\bibitem{Lebedev93}
Lebedev, V. T., Dudakov, A. D., Cser, L., Rosta, L., and Torok, G. (1993). Real space small-angle scattering device. Le Journal de Physique IV, 3(C8), C8-481.
\bibitem{Strobl14}
Strobl, M. (2014). General solution for quantitative dark-field contrast imaging with grating interferometers. Scientific reports, 4, 7243.
\bibitem{Rekveldt05}
Rekveldt, M. T., Plomp, J., Bouwman, W. G., Kraan, W. H., Grigoriev, S., and Blaauw, M. (2005). Spin-echo small angle neutron scattering in Delft. Review of Scientific Instruments, 76(3), 033901.
\end{thebibliography}
\end{document}